\documentclass[
aps, 
preprint, 
nofootinbib, 
superscriptaddress, 
prapplied,
longbibliography
]{revtex4-1}

\usepackage{amsmath}
\usepackage{amssymb}
\usepackage{graphicx}
\usepackage{dcolumn}
\usepackage{bm}
\usepackage{hyperref}
\hypersetup{
colorlinks=true,
linkcolor=blue,
citecolor=blue
}

\usepackage{float}
\usepackage{subfigure}
\usepackage{caption}
\usepackage{indentfirst}
\usepackage{mathrsfs}
\usepackage{amsfonts}

\newcommand{\ket}[1]{|{#1}\rangle}

\usepackage{color}
\definecolor{oldtxtcolor}{rgb}{0.00, 0.0, 0.5}
\definecolor{newtxtcolor}{rgb}{0.00, 0.3867, 0.00}
\definecolor{newtxtcolor}{rgb}{0.00, 0.0, 1}
\definecolor{newtxtcolor}{rgb}{0.00, 0.00, 1.00}
\definecolor{oldtxtcolor}{rgb}{1.00, 0.0, 0.00}

\def\verX{12}
\def\verO{1}
\def\verN{2}
\def\verON{12}

\ifx\verX\verO
 \newcommand { \oldtxt }[1] {{\color{oldtxtcolor}{#1}}}
 \newcommand { \newtxt }[1] {}
\fi
\ifx\verX\verN
 \newcommand { \oldtxt }[1] {}
 \newcommand { \newtxt }[1] {{\color{newtxtcolor}{#1}}}
\fi
\ifx\verX\verON
 \newcommand { \oldtxt }[1] {{\color{oldtxtcolor}{#1}}}
 \newcommand { \newtxt }[1] {{\color{newtxtcolor}{#1}}}
\fi

\begin{document}



\title{Ghost imaging with non-Gaussian quantum light}


\author{Dongyu Liu$^*$}
\affiliation{State Key Laboratory for Mesoscopic Physics and Collaborative Innovation Center of Quantum Matter, School of Physics, Peking University, Beijing 10087, China}
\author{Mingsheng Tian}
\email{These authors contributed equally to this work.}
\affiliation{State Key Laboratory for Mesoscopic Physics and Collaborative Innovation Center of Quantum Matter, School of Physics, Peking University, Beijing 10087, China}
\author{Shuheng Liu}
\affiliation{State Key Laboratory for Mesoscopic Physics and Collaborative Innovation Center of Quantum Matter, School of Physics, Peking University, Beijing 10087, China}
\author{Xiaolong Dong}
\affiliation{State Key Laboratory for Mesoscopic Physics and Collaborative Innovation Center of Quantum Matter, School of Physics, Peking University, Beijing 10087, China}
\author{Jiajie Guo}
\affiliation{State Key Laboratory for Mesoscopic Physics and Collaborative Innovation Center of Quantum Matter, School of Physics, Peking University, Beijing 10087, China}
\author{Qiongyi He}
\email{qiongyihe@pku.edu.cn}
\affiliation{State Key Laboratory for Mesoscopic Physics and Collaborative Innovation Center of Quantum Matter, School of Physics, Peking University, Beijing 10087, China}
\affiliation{{Collaborative Innovation Center of Extreme Optics, Shanxi University, Taiyuan, Shanxi 030006, China}}
\author{Haitan Xu}
\email{xuht@sustech.edu.cn}
\affiliation{Shenzhen Institute for Quantum Science and Engineering, Southern University of Science and Technology, Shenzhen 518055, China}
\affiliation{School of Physical Sciences, University of Science and Technology of China, Hefei 230026, China}
\author{Zheng Li}
\email{zheng.li@pku.edu.cn}
\affiliation{State Key Laboratory for Mesoscopic Physics and Collaborative Innovation Center of Quantum Matter, School of Physics, Peking University, Beijing 10087, China}
\affiliation{{Collaborative Innovation Center of Extreme Optics, Shanxi University, Taiyuan, Shanxi 030006, China}}
\affiliation{Peking University Yangtze Delta Institute of Optoelectronics, Nantong, China}



\date{\today}







\begin{abstract}
Non-local point-to-point correlations between two photons have been used to produce ``ghost" images without placing the camera towards the object. 
Here we theoretically demonstrated and analyzed the advantage of non-Gaussian quantum light in improving the image quality of ghost imaging system over traditional Gaussian light source.
For any squeezing degree, the signal-to-noise ratio (SNR) of the ghost image can be enhanced by the non-Gaussian operations of photon addition and subtraction on the two-mode squeezed light source. We find striking evidence that using non-Gaussian coherent operations, the SNR can be promoted to a high level even within the extremely weak squeezing regime.
The resulting insight provides 
experimental recipes of quantum imaging using non-Gaussian light for illumination.
\end{abstract}

\maketitle

\section{Introduction} 
Quantum light is the basis for developing technologies of quantum imaging \cite{Lugiato_2002,PhysRevLett.83.1763,PhysRevA.77.053807,Delaubert_2008,nphoton.2010.29,PhysRevA.83.033811,PhysRevA.77.043832,PhysRevLett.92.233601}, super-resolution\cite{PhysRevLett.85.2733,PhysRevA.79.013827,PhysRevA.67.033812,PhysRevA.77.012324}, etc. 
Especially, ghost imaging \cite{book:817050} is a representative technique based on the second-order correlation of quantum light  \cite{Klyshko_1988,PhysRevA.52.R3429}, where entangled photon pairs are used as the light source. One is able to obtain the image of an object entirely relying on the correlation between the entangled photon pairs, without directly viewing the object with a spatially resolving camera. As conceptual and practical interests are attracted to ghost imaging, the theoretical description \cite{ PhysRevA.83.063807,PhysRevA.77.043832,PhysRevA.79.023833,PhysRevLett.94.063601,PhysRevLett.99.133603,cite-key,PhysRevLett.99.133603} and improvement \cite{PhysRevA.83.063807,PhysRevLett.104.253603,doi:10.1063/1.3567931} of this technique has been studied.
The signal-to-noise ratio (SNR), one of the main parameters quantifying the quality of a ghost imaging system, characterizes how well the image is distinguished from the background~\cite{PhysRevA.83.063807}.
The two-mode squeezed state (TMSS), produced by spontaneous parametric down-conversion (SPDC) \cite{book:166611,1263776,PhysRevA.10.1874,PhysRevA.31.3093,doi:10.1080/09500348714550781}, is commonly used as the quantum light source in ghost imaging \cite{PhysRevA.52.R3429}.
It was shown that the SNR of ghost imaging with TMSS is a monotonically increasing function of the squeezing parameter of the state, associated with the degree of entanglement.
The advantage of ghost imaging with TMSS has also been shown over ghost imaging with thermal light \cite{PhysRevA.83.063807}. 
However, the SNR approaches saturation with the increase of the squeezing parameter \cite{PhysRevA.83.063807}, and due to the limitation of experimental technique, it can be difficult to prepare TMSS in the strong squeezing regime. 

Here we propose an experimental scheme to improve the SNR of ghost imaging with engineered non-Gaussian quantum light by coherently adding and subtracting photon from the {signal mode of} TMSS, i.e. {$(t\hat{a}_{{\rm s}} + r\hat{a}_{{\rm s}}^{\dagger})\ket{\mathrm{TMSS}}$}.
While the two-mode Gaussian states are frequently used as entangled photon source, it has been demonstrated that non-Gaussian operations of photon addition and subtraction~\cite{PhysRevA.84.012302,Zavatta660,Zavatta07:1890} can enhance the performance of various applications via distillation of the entangled states.
Although non-Gaussian light has proven to introduce remarkable advantage to various applications in quantum information processing, such as secure quantum communications, quantum computing and continuous variable quantum teleportation~\cite{Nha19:49,Liao18:023015,Treps20:144,PhysRevA.84.012302}, its application beyond the area of quantum information, such as quantum imaging, remains largely unexplored.
We calculate the SNR of the corresponding ghost imaging system with coherent operation {($t\hat{a}_{{\rm s}}+r\hat{a}_{{\rm s}}^{\dagger}$)}, and give the maximal SNR by optimizing the parameter $r$ of the coherent operation.
We show that the SNR can be significantly enhanced by coherent superposition operation in the weak squeezing regime, whereas the photon subtraction or addition operation performs better in the strong squeezing regime.
We also provide an analytical understanding on the mechanism how non-Gaussian local coherent operation can enhance the SNR of ghost imaging.

\section{Experimental scheme} 
\label{sec:experimental_scheme}
\begin{figure}[H]
\centering
\includegraphics[width = 0.65\textwidth]{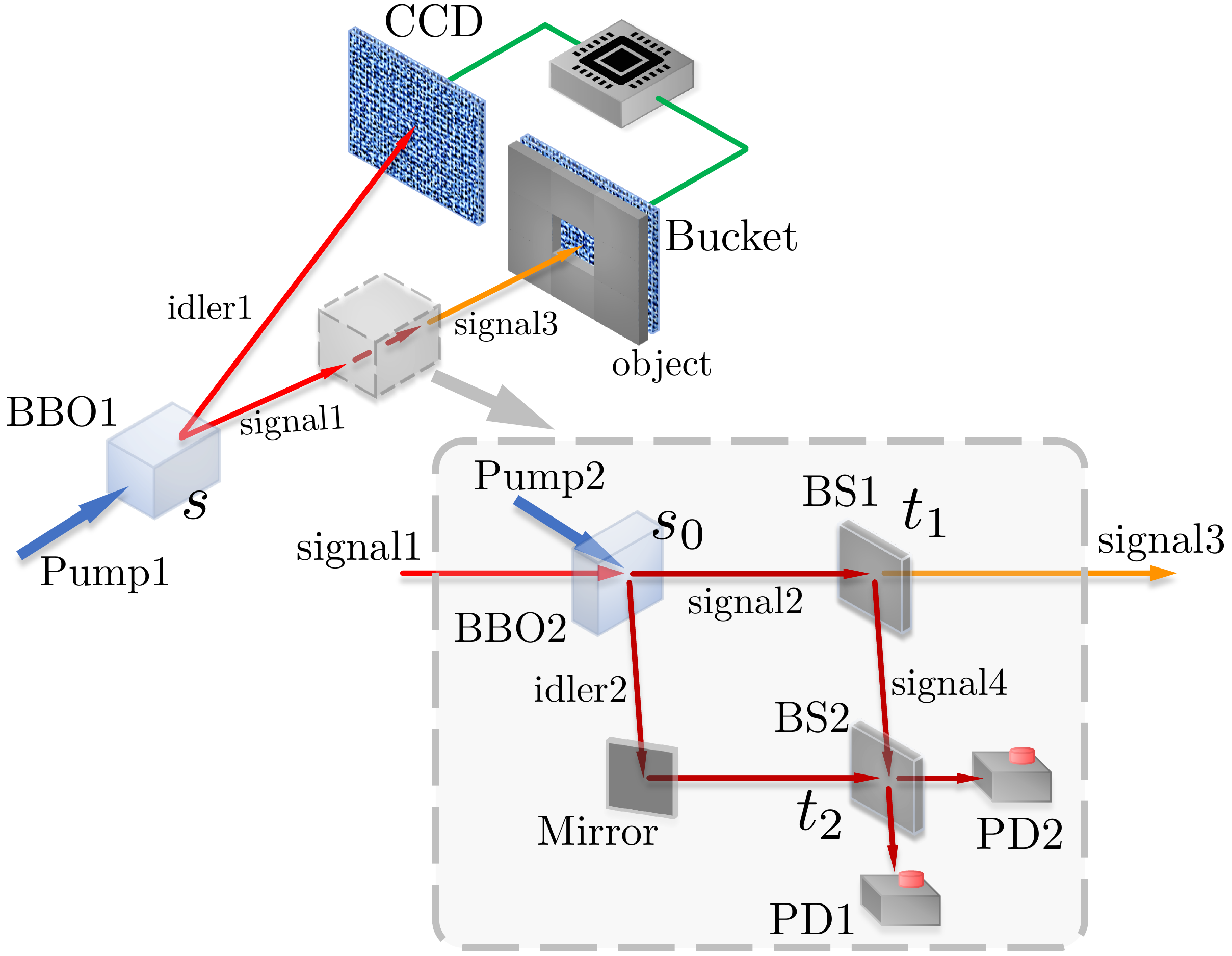}
\caption{Schematic of ghost imaging with non-Gaussian light. 
The upper part is the diagrammatic sketch of ghost imaging; 
and the lower part is the experimental scheme to implement the non-Gaussian coherent operation $t\hat{a}_{{\rm s}} + r\hat{a}_{{\rm s}}^{\dagger}$ {on the signal mode (signal1) of TMSS.} 
BS1 and BS2 are beam splitters with transmissivities $t_1$ and $t_2$, respectively.
PD1 and PD2 are photodetectors.
The detection of only a single photon at PD1 or PD2 heralds the success of a coherent operation.
{The output mode signal3 after the coherent operation and idler1 form a two-mode output state, which is then used for ghost imaging.}}
\label{fig1}
\end{figure}

The diagrammatic sketch of ghost imaging \cite{PhysRevA.52.R3429,book:817050,PhysRevA.83.063807} is shown in the upper part of FIG. \ref{fig1}. 
Entangled pairs of signal and idler photons are generated in the process of spontaneous parametric down conversion (SPDC) {in a BBO crystal (BBO1)} \cite{book:166611,1263776,PhysRevA.10.1874,doi:10.1080/09500348714550781}. 
In the regime of quantum continuous variables, it is known that the two-mode squeezed state (TMSS) of squeezing degree $s$, $\ket{\rm TMSS}= \frac{1}{\cosh s}\sum_{k=0}^{\infty}\tanh^ks{\ket{k_{\rm s}k_{\rm i}}}$, corresponds to a typical Gaussian light \cite{PhysRevA.31.3093,doi:10.1080/09500348714550781,book:2076807}.
For ghost imaging, only the signal photons are sent to the imaged object, and transmitted ones are collected by a  detector without spatial resolution (bucket detector), 
while the idler photons are sent to a spatially resolving detector (CCD). 
The image of the object is then reconstructed by measuring the function $S(x_j)$~\cite{book:817050}, where $x_j$ represents the position of the pixel $j$ of the CCD. $S$ usually has the form 
\begin{equation}
	S(x_j) = f(\langle{\mathbb{N}}_{\rm s}^pN_{\rm i}^q(x_j)\rangle),\quad p,q\geq 0,
\end{equation}
involving the correlation function $\langle{\mathbb{N}}_{\rm s}^pN_{\rm i}^q(x_j)\rangle$ of the total number of photons collected at the bucket detector, $\mathbb{N}_{\rm s}$, and at the $j\mathrm{th}$ pixel of the CCD, $N_{\rm i}$ \cite{PhysRevA.83.063807}. 
Different correlation functions can be used for ghost imaging. Here we focus on the protocol using the covariance $\mathrm{cov}(x) = \langle{\mathbb{\mathbb{N}}}_{\rm s}N_{\rm i}(x)\rangle - \langle{\mathbb{N}}_{\rm s}\rangle\langle N_{\rm i}(x)\rangle$ as the correlation function \cite{book:817050,PhysRevA.83.063807}. 

To simplify, we consider an object having only binary levels of transmission, $T=0$ and $T=1$.
The SNR of ghost image can be defined as the ratio of the mean contrast of the correlation functions inside ($T=1$) and outside ($T=0$) the object profile to the mean relative fluctuation \cite{PhysRevA.83.063807,supplementary_materials}:
\begin{equation}
	\mathrm{SNR} = \frac{|{S_{\mathrm{in}}-S_{\mathrm{out}}}|}{\sqrt{{\delta^2(S_{\mathrm{in}}-S_{\mathrm{out}})}}}.
\end{equation}

In order to improve the SNR of the ghost imaging system, we modify the Gaussian light source of ghost imaging by applying non-Gaussian operations to one of the local modes of TMSS.
In FIG. \ref{fig1}, the lower part shows the experimental scheme to implement the non-Gaussian operation, i.e. coherent operation $t\hat{a}_{{\rm s}} + r\hat{a}_{{\rm s}}^{\dagger}$ with $|t|^2+|r|^2=1$ on {the signal mode (signal1)} \cite{PhysRevA.82.053812}. 
In general, both photon subtraction ($\hat{a}$) and addition ($\hat{a}^{\dagger}$) can be implemented in quantum optics experiments. 
If an input state, e.g. {signal2 in the lower part of FIG. \ref{fig1}}, is mixed with an accessory input of a vacuum state by a beam splitter {(BS1)} with a transmissivity $t_1\simeq1$, the detection of a single photon at the accessory output port {(signal4)} heralds that a photon is subtracted from the input state \cite{PhysRevLett.92.153601}. 
On the other hand, when the input state, {e.g. signal1 in the lower part of FIG. \ref{fig1}}, is squeezed with an {idler mode in} a vacuum state in the process of parametric down conversion with a squeezing parameter $s_0 \ll 1$ {in a BBO crystal (BBO2)}, the detection of a single photon at the {idler} output port {(idler2)} heralds that a photon is added to the input state \cite{Zavatta660}. 
Combining these two operations, if the which-path information on the detected single photon is erased by an additional beam splitter (BS2) with transmissivity $t_2$ (FIG. \ref{fig1}), the operation $t\hat{a}_{{\rm s}} + r\hat{a}_{{\rm s}}^{\dagger}$ can then be conditionally implemented.
The detection of a single photon only at PD1 or PD2 heralds the success of a coherent operation.
The parameters $r$ and $t$ of the coherent operation determined by the squeezing parameter $s_0$ and the transmissivity of BS1 and BS2 are~\cite{PhysRevA.82.053812}
\begin{align}
    t = -t_2\frac{r_1}{t_1},\quad r=s_0r_2,\qquad\text{when a single photon is detected only at PD1}
    \label{transmissivity}\\
    t=r_2^*\frac{r_1}{t_1},\quad r = s_0t_2^*,\qquad \text{when a single photon is detected only at PD2}
    ,\label{reflectivity}
\end{align}
which is elaborated in the supplementary material~\cite{supplementary_materials}. 
{The parameters $t$ and $r$ are then normalized by a factor of $1/\sqrt{|t_2 r_1/t_1|^2+|s_0r_2|^2}$ for Eq.~\ref{transmissivity} and $1/\sqrt{|r_2 r_1/t_1|^2+|s_0t_2|^2}$ for Eq.~\ref{reflectivity} so that $|t|^2+|r|^2 = 1$. 
For a given BBO2 with a squeezing parameter $s_0$ (which is usually much smaller than 1), we can adjust the parameters $t_1$ and $t_2$ by choosing proper beam splitters to change the parameters $t$ and $r$, in order to implement an arbitrary coherent operation.
}When the transmissivity of BS2 is set to 1, the above operation can be reduced to photon subtraction or photon addition.

\section{Results} 
\label{sec:results_and_discussions}

\begin{figure}[H]
\centering
\subfigure[]{
\label{fig2a} 
\includegraphics[height=5.7cm]{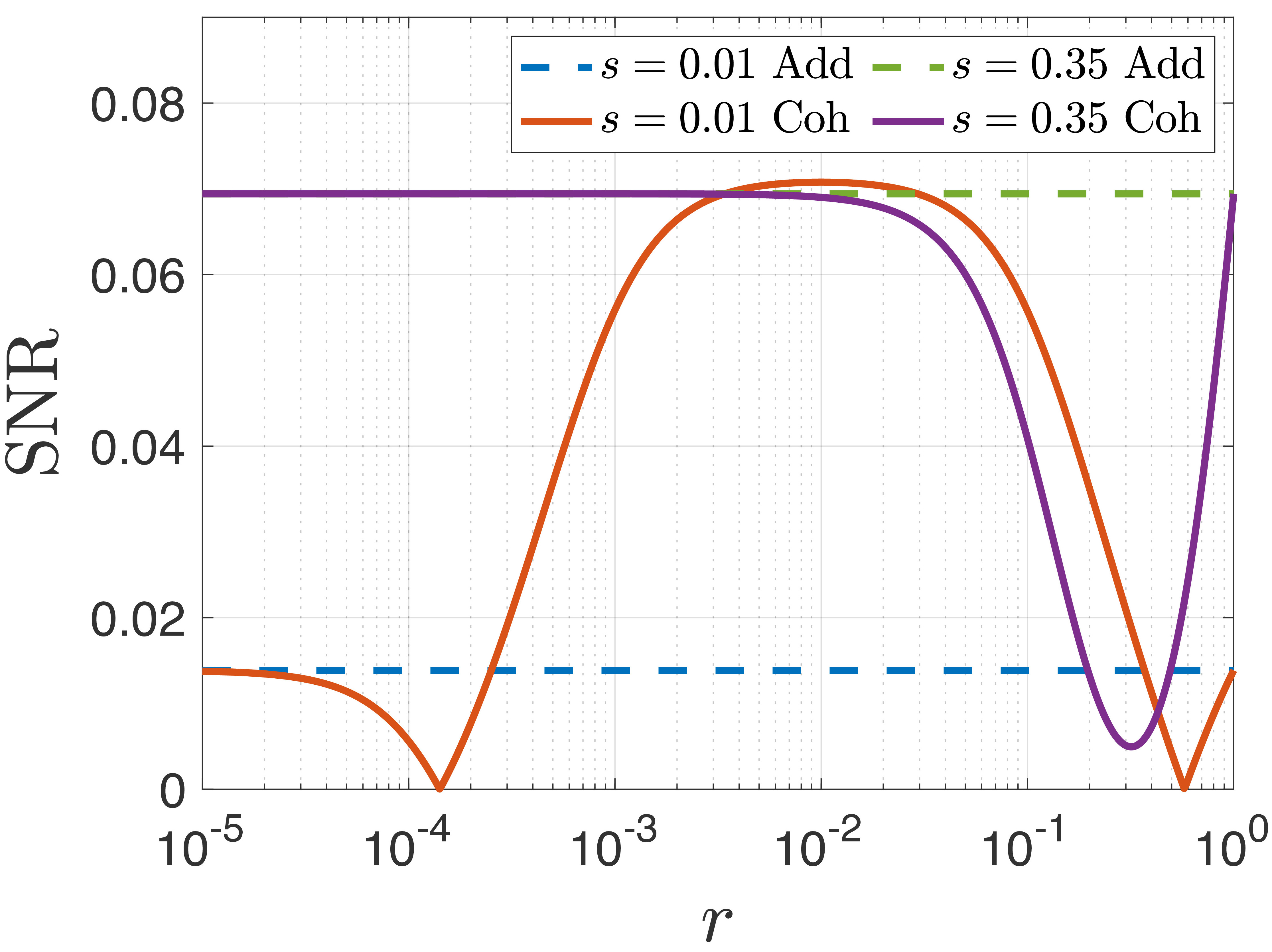}}
\subfigure[]{
\label{fig2b}
\includegraphics[height=5.8cm]{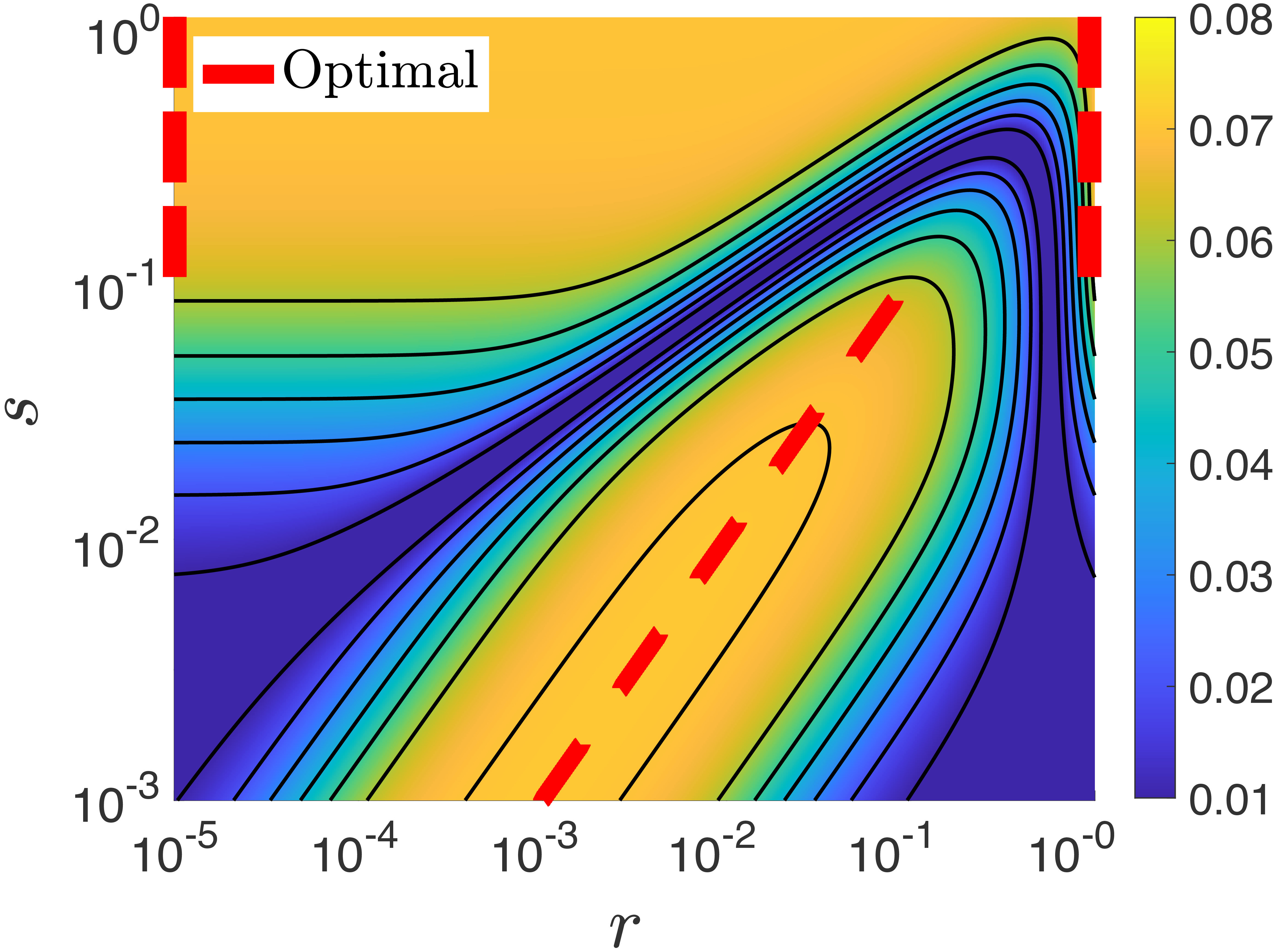}}
\centering
\caption{(a) SNR of ghost imaging as a function of the parameter $r$ of the coherent operation with (i) $s = 0.01$ for the states: $\hat{a}_{{\rm s}}^{\dagger}\ket{\mathrm{TMSS}}$ (blue dashed), $(t\hat{a}_{{\rm s}} + r\hat{a}_{{\rm s}}^{\dagger})\ket{\mathrm{TMSS}}$ (red solid), and (ii) $s = 0.35$ for the states: $\hat{a}_{{\rm s}}^{\dagger}\ket{\mathrm{TMSS}}$ (green dashed), $(t\hat{a}_{{\rm s}} + r\hat{a}_{{\rm s}}^{\dagger})\ket{\mathrm{TMSS}}$ (purple solid).
 (b) SNR of ghost imaging as a function of $s$ and $r$ for the state $(t\hat{a}_{{\rm s}} + r\hat{a}_{{\rm s}}^{\dagger})\ket{\mathrm{TMSS}}$. The SNR can be optimized by choosing proper $r$ for each $s$, and the optimal $(r,s)$ parameters are shown as red dashed line.}
\label{fig2} 
\end{figure}

We now demonstrate how the SNR of ghost imaging is enhanced with non-Gaussian operations.  
In FIG. \ref{fig2a}, we plot the SNR of the ghost imaging with non-Gaussian state of light $(t\hat{a}_{{\rm s}} + r\hat{a}_{{\rm s}}^{\dagger})\ket{\mathrm{TMSS}}$ as a function of $r$ for $s = 0.01$ and $s = 0.35$, exemplifying the effect of the coherent superposition operation on SNR. 
The photon subtraction (addition) is naturally implemented with $r = 0\ (r=1)$.
We can see that for $s=0.35$, the maximal SNR is obtained at $r = 0$ (photon subtraction) and $r = 1$ (photon addition). While for $s = 0.01$, the maximal SNR is obtained at $r \simeq s=0.01$, which is much greater than that at $r=1$ or $r=0$. 

In FIG. \ref{fig2b}, we plot the SNR of ghost imaging with $(t\hat{a}_{{\rm s}} +r\hat{a}_{{\rm s}}^{\dagger})\ket{\mathrm{TMSS}}$ in the plane of $s$ and $r$, which reveals the evolution of optimal SNR in the parameter space (red dashed line). 
In the large-squeezing regime ($s>0.09$), the optimal operation is simply photon subtraction or addition.  
Interestingly, at $s\simeq 0.09$, there appears a critical point where the optimal $(r,t)$ bifurcate abruptly to $r=0$ and $r=1$. 
Actually, in the weak squeezing regime, the optimal SNR is obtained with the coherent superposition operation at $r \simeq s$.

\begin{figure}[H]
\centering
\includegraphics[width = 0.65\textwidth]{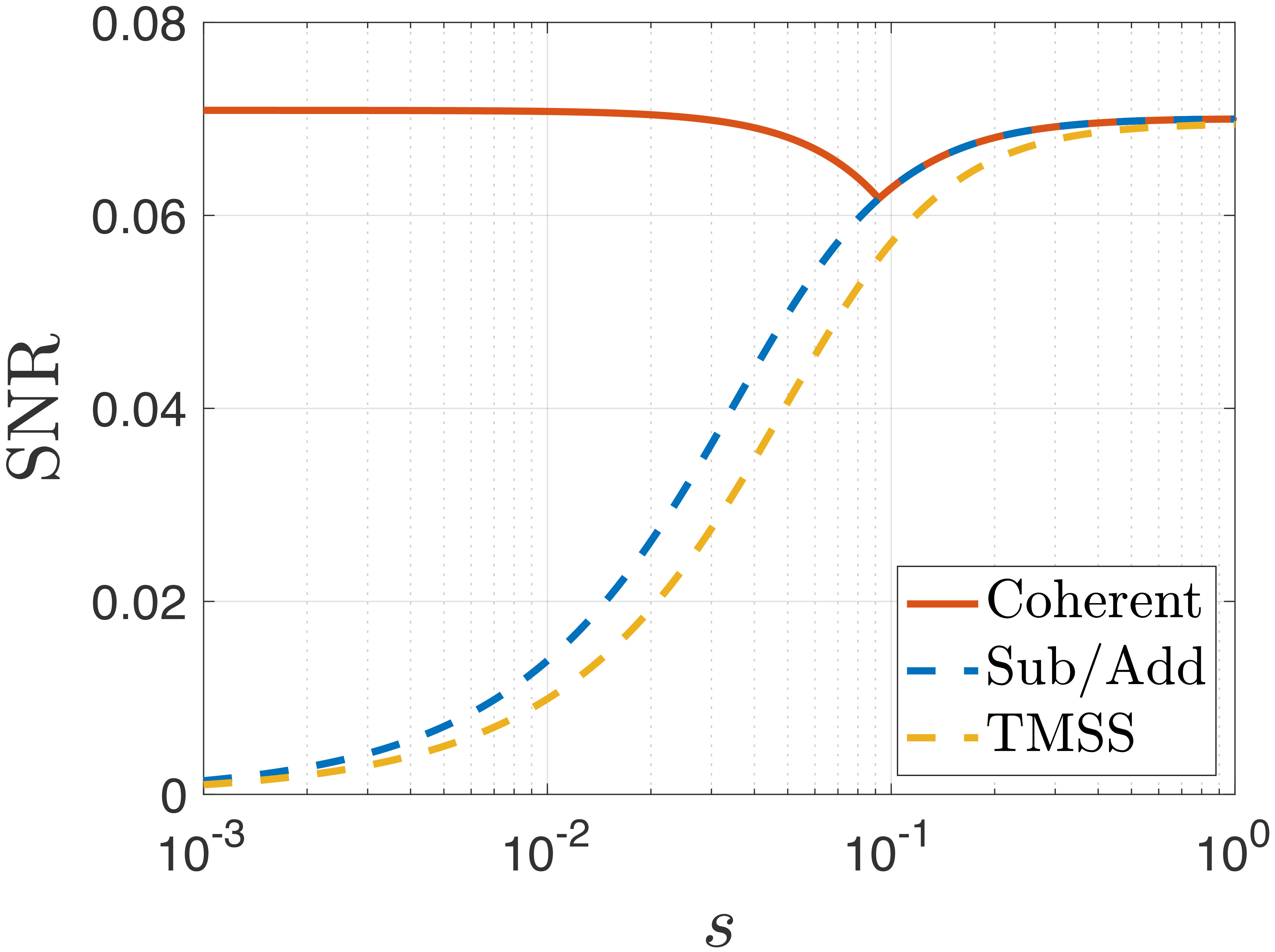}
\caption{SNR of ghost imaging as a function of $s$ for the states: $\ket{\mathrm{TMSS}}$ (yellow dashed), $\hat{a}_{{\rm s}}\ket{\mathrm{TMSS}}$ (or $\hat{a}_{{\rm s}}^{\dagger}\ket{\mathrm{TMSS}}$) (blue dashed), and $(t\hat{a}_{{\rm s}} + r\hat{a}_{{\rm s}}^{\dagger})\ket{\mathrm{TMSS}}$ (red solid). The parameter $r$ in the coherent operation is optimized for each $s$.}
\label{fig3}
\end{figure}

In FIG. \ref{fig3}, we show the optimized SNR of ghost imaging with non-Gaussian state of light  $(t\hat{a}_{{\rm s}} + r\hat{a}_{{\rm s}}^{\dagger})\ket{\mathrm{TMSS}}$ as a function of the squeezing parameter $s$, compared with those using $\hat{a}_{{\rm s}}\ket{\mathrm{TMSS}}$ (or $\hat{a}_{{\rm s}}^{\dagger}\ket{\mathrm{TMSS}}$) and the Gaussian state $\ket{\mathrm{TMSS}}$ as light source. There is a slight improvement of SNR via  photon subtraction (addition). 
However, the SNR is remarkably enhanced by the coherent operation $t\hat{a}_{{\rm s}} + r\hat{a}_{{\rm s}}^{\dagger}$ in the weak squeezing region, and reaches a maximal level even greater than that of strongly squeezed state. The turning point at $s\simeq 0.091$ comes from the abrupt change of the optimized operation from coherent operation to photon subtraction or addition. 
Given the technical difficulties in preparing TMSS with large squeezing parameter, such enhancement of SNR can be quite useful for ghost imaging with quantum light in the weak squeezing regime. 
The SNR of ghost imaging with TMSS in the experiment-friendly weak-squeezing regime can be significantly enhanced via the coherent operation. 
This circumvents the difficulty of preparing TMSS with large squeezing parameter, using the conditional implementation of non-Gaussian operation where strong squeezing process is not necessary either.

We provide an analytical understanding of the behavior of the SNR with non-Gaussian light $(t\hat{a}_{{\rm s}} + r\hat{a}_{{\rm s}}^{\dagger})\ket{\mathrm{TMSS}}$. 
Assuming the squeezing parameter $s\ll 1$ in the weak-squeezing regime, the TMSS can be approximated by 
\begin{align}
\ket{\rm TMSS} = \frac{1}{\cosh s}\sum_{k=0}^{\infty}\tanh^k s\ {\ket{k_{\rm s}k_{\rm i}}} \simeq {\ket{0_{\rm s}0_{\rm i}}} + {s\ket{1_{\rm s}1_{\rm i}}} \label{TMSS_approx}
\end{align}
The non-Gaussian coherent operation $t\hat{a}_{{\rm s}} + r\hat{a}_{{\rm s}}^{\dagger}$ on the TMSS with $|t|^2+|r|^2=1$ then generates an output state as \begin{align}
(t\hat{a}_{{\rm s}} + r\hat{a}_{{\rm s}}^{\dagger})\ket{\mathrm{TMSS}} \simeq r{\ket{1_{\rm s}0_{\rm i}}} + s(t{\ket{0_{\rm s}1_{\rm i}}} + \sqrt{2}r{\ket{2_{\rm s}1_{\rm i}}}) \label{coh_approx}
\end{align}
According to the optimization strategy in the weak-squeezing regime, the optimal SNR is obtained at $r \simeq s \ll 1$, with the output state given by 
\begin{align}
(t\hat{a}_{{\rm s}} + r\hat{a}_{{\rm s}}^{\dagger})\ket{\mathrm{TMSS}} &\simeq s{\ket{1_{\rm s}0_{\rm i}}} + s\sqrt{1-s^2}{\ket{0_{\rm s}1_{\rm i}}} + \sqrt{2}s^2{\ket{2_{\rm s}1_{\rm i}}}\notag\\
& \rightarrow\frac{1}{\sqrt{2}}({\ket{1_{\rm s}0_{\rm i}}}+{\ket{0_{\rm s}1_{\rm i}}})  \label{Bell_approx}
\end{align}
Thus we obtain a Bell state $\frac{1}{\sqrt{2}}({\ket{1_{\rm s}0_{\rm i}}}+{\ket{0_{\rm s}1_{\rm i}}})$ as an approximation of the non-Gaussian light in the weak squeezing regime.
The calculated SNR given by this Bell state is about 0.07, which coincides well with the optimal SNR of $(t\hat{a}_{{\rm s}} + r\hat{a}_{{\rm s}}^{\dagger})\ket{\rm TMSS}$  in the weak squeezing regime. 
On the other hand, the output states cannot be reduced to a Bell state by photon subtraction or addition alone, resulting in less enhancement of SNR than by the coherent $t\hat{a}_{{\rm s}} + r\hat{a}_{{\rm s}}^{\dagger}$ operation.
The optimized $(t\hat{a}_{{\rm s}} + r\hat{a}_{{\rm s}}^{\dagger})\ket{\mathrm{TMSS}}$ greatly differs from $\hat{a}_{{\rm s}}\ket{\mathrm{TMSS}}$ (or $\hat{a}_{{\rm s}}^{\dagger}\ket{\mathrm{TMSS}}$) in the weak-squeezing regime. 
As the squeezing parameter $s$ increases, the higher order terms of $s$ in Eq.~(\ref{TMSS_approx}),(\ref{coh_approx}) can no longer be neglected, and the approximation to Bell state in Eq.~(\ref{Bell_approx}) becomes invalid.
Due to the contribution of the component states related to the higher order terms of $s$ in $(t\hat{a}_{{\rm s}} + r\hat{a}_{{\rm s}}^{\dagger})\ket{\rm TMSS}$, such as ${\ket{1_{\rm s}2_{\rm i}}}$ and ${\ket{2_{\rm s}1_{\rm i}}}$, the SNR decreases relevantly. 
When $(t\hat{a}_{{\rm s}} + r\hat{a}_{{\rm s}}^{\dagger})\ket{\mathrm{TMSS}}$ no longer shows superiority to $\hat{a}_{{\rm s}}\ket{\mathrm{TMSS}}$ (or $\hat{a}_{{\rm s}}^{\dagger}\ket{\mathrm{TMSS}}$) in ghost imaging, the turning point appears, and the enhancement of SNR is then optimized by the photon subtraction or addition operation. 

\section{Conclusion} 
In summary, we have proposed an experimental scheme to improve the SNR of ghost imaging by non-Gaussian coherent operation, instead of increasing the squeezing parameter of TMSS.
We have achieved a remarkable increase of SNR of ghost imaging in the weak squeezing regime, demonstrated by numerical simulation. 
We have also given an analytical understanding of the enhancement of SNR via coherent operation, which is related to the EPR correlation.
The enhancement of SNR can be regarded as a significant advance for ghost imaging with quantum light, overcoming the limitation of SNR due to the technical difficulty in preparing states in the strong squeezing regime.
In addition, it can be considered as a feasible application of the coherent superposition operation of photon subtraction and addition.

\section{Acknowledgements}
{This work is supported by National Natural Science Foundation of China (Grant Nos. 12174009, 11974031).}
\newpage
\bibliography{citation}

\end{document}



\title{Supplemental Material\\Ghost imaging with non-Gaussian quantum light}


\author{Dongyu Liu$^*$}
\affiliation{State Key Laboratory for Mesoscopic Physics and Collaborative Innovation Center of Quantum Matter, School of Physics, Peking University, Beijing 10087, China}
\author{Mingsheng Tian}
\email{These authors contributed equally to this work.}
\affiliation{State Key Laboratory for Mesoscopic Physics and Collaborative Innovation Center of Quantum Matter, School of Physics, Peking University, Beijing 10087, China}
\author{Shuheng Liu}
\affiliation{State Key Laboratory for Mesoscopic Physics and Collaborative Innovation Center of Quantum Matter, School of Physics, Peking University, Beijing 10087, China}
\author{Xiaolong Dong}
\affiliation{State Key Laboratory for Mesoscopic Physics and Collaborative Innovation Center of Quantum Matter, School of Physics, Peking University, Beijing 10087, China}
\author{Jiajie Guo}
\affiliation{State Key Laboratory for Mesoscopic Physics and Collaborative Innovation Center of Quantum Matter, School of Physics, Peking University, Beijing 10087, China}
\author{Qiongyi He}
\email{qiongyihe@pku.edu.cn}
\affiliation{State Key Laboratory for Mesoscopic Physics and Collaborative Innovation Center of Quantum Matter, School of Physics, Peking University, Beijing 10087, China}
\affiliation{{Collaborative Innovation Center of Extreme Optics, Shanxi University, Taiyuan, Shanxi 030006, China}}
\author{Haitan Xu}
\email{xuht@sustech.edu.cn}
\affiliation{Shenzhen Institute for Quantum Science and Engineering, Southern University of Science and Technology, Shenzhen 518055, China}
\affiliation{School of Physical Sciences, University of Science and Technology of China, Hefei 230026, China}
\author{Zheng Li}
\email{zheng.li@pku.edu.cn}
\affiliation{State Key Laboratory for Mesoscopic Physics and Collaborative Innovation Center of Quantum Matter, School of Physics, Peking University, Beijing 10087, China}
\affiliation{{Collaborative Innovation Center of Extreme Optics, Shanxi University, Taiyuan, Shanxi 030006, China}}
\affiliation{Peking University Yangtze Delta Institute of Optoelectronics, Nantong, China}


\date{\today}






\maketitle

\section{Theory of signal-to-noise ratio in ghost imaging with non-Gaussian quantum light} 
\label{sec:theory_of_signal_to_noise_ratio_in_ghost_imaging_with_non_Gaussian_light}

In ghost imaging, the image of an object is generally retrieved by measuring the function $S(x_j)$, in which $x_j$ represents the position of the $j\textrm{th}$ pixel of the spatially resolving detector (CCD). 
$S$ usually has the form
\begin{equation}
	S(x_j) = f(E[\mathbb{N}_{\rm s}^pN_{\rm i}^q]),\quad p,q\geq 0,
\end{equation}
involving the correlation function $E[\mathbb{N}_{\rm s}^pN_{\rm i}^p]$ of the total number of photons collected at the bucket detector, $\mathbb{N}_{\rm s}$, and at the $j\textrm{th}$ pixel of the CCD, $N_{\rm i}(x_j)$, respectively \cite{PhysRevA.83.063807}. 
Here, $E[X]$ represents the average of $X$ over the measuring time in an experimental scheme, which is equivalent to the expectation value in quantum mechanics, denoted by $\qav{X}$. To simplify, we consider an imaged object which has only two levels of transmission, $T =0$ and $T =1$. 

The SNR of ghost image can be defined as the ratio of the mean contrast of the correlation function inside ($T = 1,\ S_{\mathrm{in}}$) and outside ($T = 0,\ S_{\mathrm{out}}$) of the object profile, to its standard deviation \cite{PhysRevA.83.063807}:
\begin{equation}
	\mathrm{SNR} = \frac{|\qav{S_{\mathrm{in}}-S_{\mathrm{out}}}|}{\sqrt{\qav{\delta^2(S_{\mathrm{in}}-S_{\mathrm{out}})}}}.
\end{equation}

 There are several ghost imaging protocols with different correlation functions\cite{PhysRevA.83.063807}. Here we consider the protocol based on using the covariance as correlation function, 
\begin{equation}
	S(x) \equiv \mathrm{cov}(x) = \qav{[\mathbb{N}_{\rm s}-\qav{\mathbb{N}_{\rm s}}][{N}_{\rm i}(x)-\qav{{N}_{\rm i}(x)}]}.
\end{equation}

In the covariance protocol, SNR can be expressed by the following first- to fourth-order moments:
$\qav{\mathbb{N}_{\rm s}},\ \qav{N_{\rm i}},\ \qav{\mathbb{N}_{\rm s}^2},\ \qav{N_{\rm i}^2},\ \qav{\mathbb{N}_{\rm s}N_{\rm i}},\ \qav{\mathbb{N}_{\rm s}^2N_{\rm i}},\ \qav{\mathbb{N}_{\rm s}N_{\rm i}^2},\ \qav{\mathbb{N}_{\rm s}^2N_{\rm i}^2}$. 
And all these terms can be further expressed by photon-number statistic, $\qav{\hat{n}_{\rm s}^{p}\hat{n}_{\rm i}^{q}}$, where $\hat{n}_\alpha = \hat{a}_\alpha^{\dagger}\hat{a}_\alpha,\ \alpha = {\rm s},{\rm i}$. Detailed information as well as calculation can be found in \cite{PhysRevA.83.063807}.
Hence, the calculation of SNR of ghost imaging with quantum light is reduced to the calculation of photon-number statistic as shown below.

The two-mode squeezed state (TMSS), produced by spontaneous parametric down-conversion (SPDC), is commonly used as a quantum light source in ghost imaging. 
In the regime of quantum continuous variables,
\begin{equation}
	\ket{\Psi_{\mathrm{TMSS}}}=\frac{1}{\cosh s}\sum_{k=0}^{\infty} \tanh^k s\ \ket{k_{\rm s}k_{\rm i}}.
\end{equation}

In order to enhance the SNR in ghost imaging with quantum light, we implement the coherent operation\cite{PhysRevA.84.012302,PhysRevA.82.053812}, $t\hat{a}_{\rm s}+r\hat{a}_{\rm s}^{\dagger}$, where $|t|^2+|r|^2=1$, on the signal mode of TMSS, and obtain the (unnormalized) output state as \cite{PhysRevA.84.012302}
\begin{equation}
	\ket{\psi_{\pm}} = (t\hat{a}_{\rm s}+r\hat{a}_{\rm s}^{\dagger})\ket{\Psi_{\mathrm{TMSS}}} = \frac{1}{\cosh s}\sum_{k=0}^{\infty} \tanh^k s(t\hat{a}_{\rm s}+r\hat{a}_{\rm s}^{\dagger})\ket{k_{\rm s}k_{\rm i}}.
\end{equation}

One can calculate the photon-number statistic of $(t\hat{a}_{\rm s}+r\hat{a}_{\rm s}^{\dagger})\ket{\Psi_{\mathrm{TMSS}}}$, 
\begin{equation}
	\qav{\hat{n}_{\rm s}^p\hat{n}_{\rm i}^q} = \frac{\braket{\psi_{\pm}}{\hat{n}_{\rm s}^p\hat{n}_{\rm i}^q}{\psi_{\pm}}}{\inn{\psi_{\pm}}{\psi_{\pm}}} = \frac{\braket{\Psi_{\mathrm{TMSS}}}{(t\hat{a}_{\rm s}^{\dagger}+r\hat{a}_{\rm s})\hat{a}_{\rm s}^{\dagger p}\hat{a}_{\rm s}^p\hat{a}_{\rm i}^{\dagger q}\hat{a}_{\rm i}^q(t\hat{a}_{\rm s}+r\hat{a}_{\rm s}^{\dagger})}{\Psi_{\mathrm{TMSS}}}}{\braket{\Psi_{\mathrm{TMSS}}}{(t\hat{a}_{\rm s}^{\dagger}+r\hat{a}_{\rm s})(t\hat{a}_{\rm s}+r\hat{a}_{\rm s}^{\dagger})}{\Psi_{\mathrm{TMSS}}}}.
\end{equation}

The analytic expressions of photon-number statistic are given as follows: 
\begin{align}
\left\langle n_{\rm s}\right\rangle &= \frac{\left(6 r^2-4\right) \cosh 2 s-2 r^2+\cosh 4 s+3}{4 \left(r^2+\sinh ^2s\right)}\\
\left\langle n_{\rm i}\right\rangle &= \frac{\sinh ^2s \left(r^2+\cosh 2s\right)}{r^2+\sinh ^2s}\\
\left\langle n_{\rm s}^2\right\rangle &=\frac{\sinh ^2s \left[4 \left(5 r^2-3\right) \cosh 2s-4 r^2+3 \cosh 4s+9\right]}{4 \left(r^2+\sinh ^2s\right)}+ \left\langle n_{\rm s}\right\rangle\\
\left\langle n_{\rm i}^2\right\rangle &=\frac{\sinh ^4s \left(2 r^2+3 \cosh 2s+1\right)}{r^2+\sinh ^2s}+\left\langle n_{\rm i}\right\rangle\\
\left\langle n_{\rm s} n_{\rm i}\right\rangle &= \sinh ^2s (3 \cosh 2s+1)\\
\left\langle n_{\rm s}^2 n_{\rm i}\right\rangle  &= \frac{\sinh ^2s \left[\left(9-4 r^2\right) \cosh 2s+3 \left(5 r^2-3\right) \cosh 4s+5 r^2+3 \cosh 6s-3\right]}{4 \left(r^2+\sinh ^2s\right)}+\left\langle n_{\rm s} n_{\rm i}\right\rangle\\
\left\langle n_{\rm s} n_{\rm i}^2\right\rangle &= \frac{\sinh ^4s \left(9 r^2 \cosh 2s+5 r^2+3 \cosh 4s+1\right)}{r^2+\sinh ^2s} +\left\langle n_{\rm s} n_{\rm i}\right\rangle\\
\left\langle n_{\rm s} n_{\rm i}\right\rangle &=3 \sinh ^4s (4 \cosh 2s+5 \cosh 4s+3)+\left\langle n_{\rm s}^2 n_{\rm i}\right\rangle +\left\langle n_{\rm s} n_{\rm i}^2\right\rangle -\left\langle n_{\rm s}^2 n_{\rm i}^2\right\rangle 
\end{align}

\section{The experimental scheme to implement the coherent operation} 
\label{sec:the_experimental_scheme_to_implement_the_coherent_operation}
\begin{figure}[h]
\centering
\includegraphics[width=0.5\textwidth]{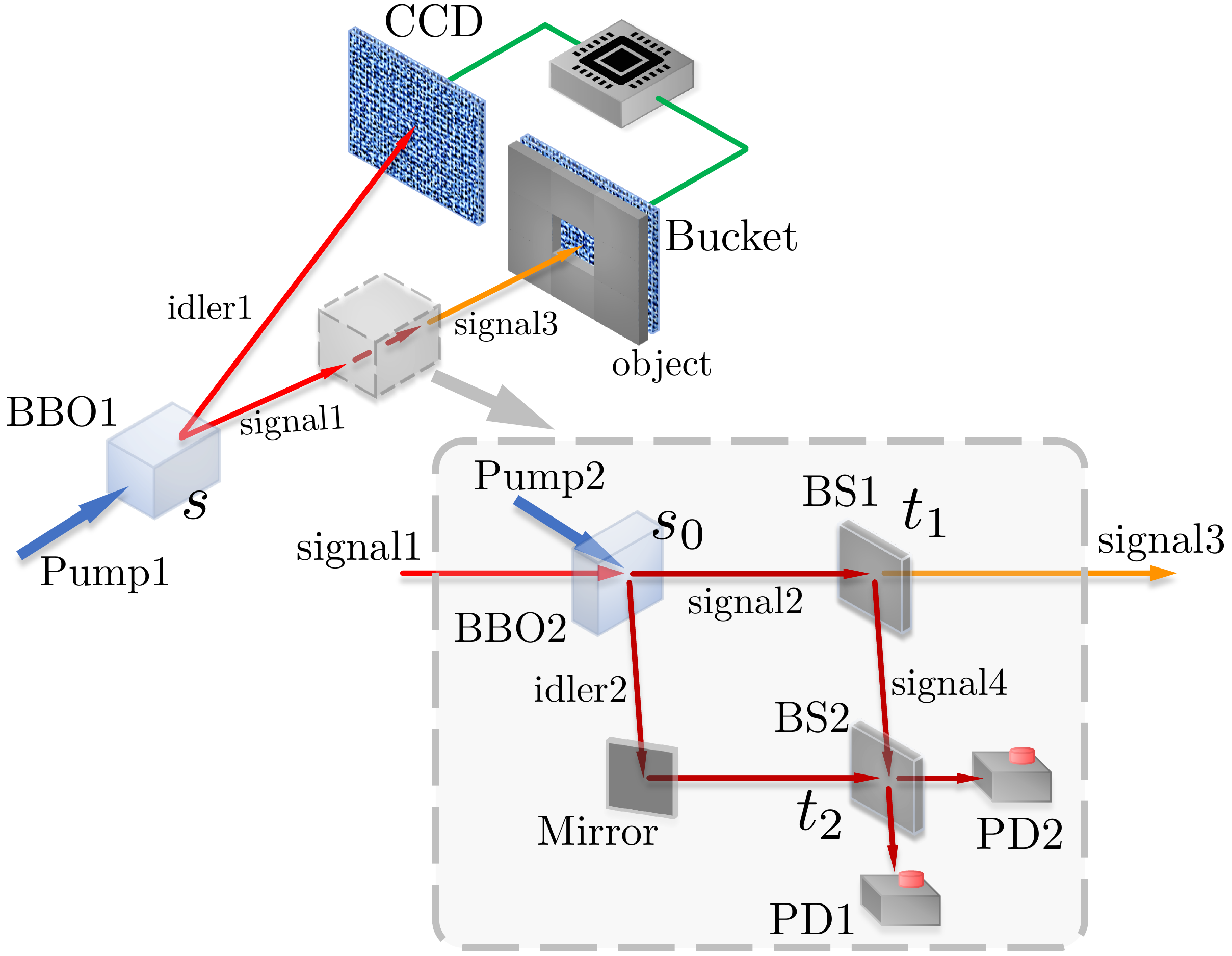}
\caption{Experimental scheme to implement the coherent operation $t\hat{a} + r \hat{a}^{\dagger}$ on an arbitrary state $\ket{\psi}$. 
BS1 and BS2 are beam splitters with transmissivities $t_1$ and $t_2$, respectively. 
PD1 and PD2: photo detectors. 
The coherent operation is successfully achieved under the detection of a single photon only at PD1 or PD2.} 
\label{scheme}
\end{figure}

In FIG. \ref{scheme}, an arbitrary state $\ket{\psi}$ is injected to the signal mode of a nondegenerate parametric amplifier with small squeezing parameter $s_0\ll1$, which acts as~\cite{PhysRevA.82.053812}
\begin{align}
	\exp(s_0\hat{a}^{\dagger}\hat{c}^{\dagger}-s_0\hat{a}\hat{c})\ket{\psi}_a\ket{0}_c &= [1+(s_0\hat{a}^{\dagger}\hat{c}^{\dagger}-s_0\hat{a}\hat{c})+\cdots]\ket{\psi}_a\ket{0}_c\notag\\
	&\simeq (1+s_0\hat{a}^{\dagger}\hat{c}^{\dagger})\ket{\psi}_a\ket{0}_c
\end{align}
Then, the beams splitter (BS1) with transmissivity $t_1\simeq1$ acts on the state as
\begin{equation}
    \left(1-\frac{r_1}{t_1}\hat{a}\hat{b}^{\dagger}\right)(1+s_0\hat{a}^{\dagger}\hat{c}^{\dagger})\ket{\psi}_a\ket{0}_b\ket{0}_c
\end{equation}

Finally, beam splitter BS2 with transmissivity $t_2$ acts on the two modes, $b$ and $c$. 
The following substitutions should be made
\begin{align}
\hat{b} \rightarrow t_2^*\hat{b} - r_2\hat{c}\notag\\
\hat{c} \rightarrow t_2\hat{c} + r_2^*\hat{b}
\end{align}
Thus, the final output state is given by 
\begin{align}
\left[ \right.1 &\left.-\frac{r_1}{t_1}\hat{a}\left(t_2\hat{b}^{\dagger}-r_2^*\hat{c}^{\dagger}\right)+ s_0\hat{a}^{\dagger}\left(t_2^*\hat{c}^{\dagger}+r_2\hat{b}^{\dagger}\right)\right.\notag\\
&\left.-\frac{r_1}{t_1}s_0\hat{a}\hat{a}^{\dagger}\left(t_2\hat{b}^{\dagger}-r_2^*\hat{c}^{\dagger}\right)\left(t_2^*\hat{c}^{\dagger}+r_2\hat{b}^{\dagger}\right)\right]\ket{\psi}_a\ket{0}_b\ket{0}_c
\label{final_output}
\end{align}

With the detection of single photon at PD1 and no photon at PD2, we see from Eq. (\ref{final_output}) that the state collapses to 
\begin{align}
	\ket{\psi}_{\rm out} &= \left(-t_2\frac{r_1}{t_1}\hat{a}+s_0r_2\hat{a}^{\dagger}\right)\ket{\psi}_a\notag\\
	& = (t\hat{a}+r\hat{a}^{\dagger})\ket{\psi}_a
\end{align}
where $t = -t_2\frac{r_1}{t_1},r=s_0r_2$. 
Similarly, with the detection of single photon at PD2 and no photon at PD1, we have the same result, $\ket{\psi}_{\rm out} = (t\hat{a}+r\hat{a}^{\dagger})\ket{\psi}_a$, where $t=r_2^*\frac{r_1}{t_1},r = s_0t_2^*$.
\bibliography{citation}